\documentclass[aip,graphicx]{revtex4-1}

\draft 

\begin{document}

\title{Tetrads in $SU(3) \times SU(2) \times U(1)$ Yang-Mills geometrodynamics} 

\author{Alcides Garat}
\email[]{garat.alcides@gmail.com}
\affiliation{1. Instituto de F\'{\i}sica, Facultad de Ciencias, Igu\'a 4225, esq. Mataojo, Montevideo, Uruguay.}
\date{December 24th, 2011}

\begin{abstract}

The relationship between gauge and gravity amounts to understanding underlying new geometrical local structures. These structures are new tetrads specially devised for Yang-Mills theories, Abelian and Non-Abelian in four-dimensional Lorentzian curved spacetimes. In the present manuscript a new tetrad is introduced for the Yang-Mills $SU(3) \times SU(2) \times U(1)$ formulation. These new tetrads establish a link between local groups of gauge transformations and local groups of spacetime transformations that we previously called LB1 and LB2. New theorems are proved regarding isomorphisms between local internal $SU(3) \times SU(2) \times U(1)$ groups and local tensor products of spacetime LB1 and LB2 groups of transformations. These new tetrads define at every point in spacetime two orthogonal planes that we called blades or planes one and two. These are the local planes of covariant diagonalization of the stress-energy tensor. These tetrads are gauge dependent. Tetrad local gauge transformations leave the tetrads inside the local original planes without leaving them. These local tetrad gauge transformations enable the possibility to connect local gauge groups Abelian or non-Abelian with local groups of tetrad transformations. On the local plane one, the Abelian group $U(1)$ of gauge transformations was already proved to be isomorphic to the tetrad local group of transformations LB1, for example. LB1 is $SO(1,1)$ plus two different kinds of discrete transformations. On the local orthogonal plane two $U(1)$ is isomorphic to LB2 which is just $SO(2)$. That is, we proved that LB1 is isomorphic to $SO(2)$ which is a remarkable result since a non-compact group plus two discrete transformations is isomorphic to a compact group. These new tetrads have displayed manifestly and non-trivially the coupling between Yang-Mills fields and gravity. The new tetrads and the stress-energy tensor allow for the introduction of three new local gauge invariant objects. Using these new gauge invariant objects and in addition a new general local duality transformation, a new algorithm for the gauge invariant diagonalization of the Yang-Mills stress-energy tensor is developed as an application. This is a paper about grand Standard Model gauge theories - General Relativity gravity unification and grand group unification in four-dimensional curved Lorentzian spacetimes.
\end{abstract}

\keywords{new tetrads; new groups; new group isomorphisms; gravity-Yang-Mills classical unification}

\pacs{02.40.ky; 04.20.-q; 11.15.-q; 04.20.Cv \\ MSC2010: 20F65; 70s15; 70G65; 70G45; 53c50}

\maketitle 

\section{Introduction}
\label{intro}

The geometrization of gauge theories is the focus of our attention. We want to develop suitable tools in order to understand the nature of gravitational fields in the presence of local Abelian and non-Abelian Yang-Mills fields. In addition we find fundamental results in group theory and new techniques in dealing with gauge invariant diagonalization of stress-energy tensors \cite{A}$^{-}$\cite{MW}. In the non-Abelian case through a new kind of duality transformations. We also establish explicitly the relationship between gauge fields and gravity through non-trivial new tetrads. Differential equations will also be simplified.  Other future applications are also possible. For instance, the study of the kinematics in these spacetimes \cite{RG}$^{-}$\cite{EJKS}, that is, the search for a ``connection'' to the theory of embeddings, time slicings \cite{KK}$^{-}$\cite{FAEP}, the initial value formulation \cite{JWY1}$^{-}$\cite{CMDWYCB}, the Cauchy evolution \cite{AEO}$^{-}$\cite{LSJWY}, etc. These tools have been investigated in previous works for the electromagnetic Abelian $U(1)$ and the $SU(2) \times U(1)$ non-Abelian cases \cite{A}$^{,}$\cite{A2}$^{,}$\cite{A3}. New tetrads were introduced in four-dimensional Lorentzian curved space-times. These new tetrads on one hand allow to translate all the standard gauge theories into a new mathematical language, and on the other hand produce substantial simplification in field equations, the diagonalization of stress-energy tensors in a gauge invariant way, etc. These same new tetrads allowed to prove theorems about isomorphisms between local gauge groups of transformations and the local group of transformations LB1, LB2 and their tensor products. These last results amount to prove that the no-go theorems from the mid-sixties \cite{SWNG}$^{,}$\cite{LORNG}$^{,}$\cite{CMNG} were developed under wrong hypotheses. It was thought that local groups of gauge ``internal'' transformations do not act on spacetime objects. The internal local groups of transformations do not act upon spacetime variables. In the aforementioned works  \cite{A}$^{,}$\cite{A2}$^{,}$\cite{A3} isomorphisms were proved between local groups of gauge transformations and the local groups LB1 and LB2. The new tetrads were the tool necessary to prove these new theorems. These theorems cancel the hypotheses of the no-go theorems. For instance, in the $U(1)$ case it was proved that at every point in a four-dimensional Lorentzian spacetime we can build a new special kind of tetrad such that two orthogonal planes of covariant diagonalization of the stress-energy tensor are defined at every tangent space. One plane spanned by a timelike and one spacelike vectors that we named blade one. The other plane orthogonal to blade one spanned by the other two spacelike vectors, blade two. Every vector on these local orthogonal planes is an eigenvector to the Einstein-Maxwell stress-energy tensor. The construction of these tetrad vectors involves extremal fields on one hand in order to build the tetrad  ``skeleton'', and on the other hand two vector fields,  the ``gauge vectors''. It happens that we can prove that if we choose the gauge vectors to be either of the two electromagnetic potentials \cite{A} that exist in the Abelian case, two theorems can be proved for this Abelian case. The first theorem states that the local group of electromagnetic gauge transformations is isomorphic to the local group of Lorentz proper tetrad transformations on blade one plus two kinds of discrete transformations, one of them is not a Lorentz transformation, on this same plane, the group LB1.  The second theorem states that the local group of electromagnetic gauge transformations is isomorphic to the local group of Lorentz tetrad rotations $SO(2)$ on blade two, the group LB2. This last result is not new, what is new is that LB1 is isomorphic to LB2, or $SO(2)$. That is, just to give an example, to every proper transformation on blade one corresponds a space rotation on blade two. Besides, since the local $U(1)$ transformations are isomorphic to the local Lorentz LB1 and LB2 transformations on special unique orthogonal planes of diagonalization of the Einstein-Maxwell stress-energy tensor, then the LB1 and LB2 groups do not commute in general with any local Lorentz transformation. Because all but one of the discrete transformations in LB1 which is not a Lorentz transformation, are indeed Lorentz transformations, and the Lorentz transformations on planes one and two do not commute in general with any possible local Lorentz transformation on any other local plane. Tetrad Lorentz transformations on a special plane, do not commute in general with tetrad Lorentz transformations on a different plane at the same point or tangent space. Therefore, the hypotheses of the no-go theorems do not apply in general. Furthermore, for local $SU(2)$ gauge transformations, similar theorems were proved involving again special tetrads and tensor products of LB1 and LB2 local groups \cite{A2}$^{,}$\cite{A3}.  It is really important that we make a clarification on an issue that might give rise to confusion.
In these theorems for local $SU(2)$ gauge transformations \cite{A2}$^{,}$\cite{A3} we are considering three copies of the same spacetime, and a different tetrad at the same point in each spacetime copy. These three tetrads have a mutually similar extremal field-gauge vector structure. They are normalized and a choice of gauge vector has been made. But they are not the same. They could be Lorentz transformed into each other, under non-trivial Lorentz spatial rotations, for instance. We know from manuscript \cite{A2} that a local Lorentz transformation of a tetrad itself with an extremal field-gauge vector structure transforms into another tetrad with a similar extremal field-gauge vector structure, even though the skeletons will not be the same, of course. Therefore, we are considering three local tetrads at the same spacetime point which are not the same for different copies. This is what we mean by three LB1 or LB2 groups under tensor product. Now, from an operational point of view what we also mean by these theorems is that we are able to reconstruct the original local $SU(2)$ gauge transformation by knowing for example, the boosts in the LB1 case, or the spatial rotations in the LB2 case, for the tetrad local transformations at the point under consideration. By knowing the local Lorentz transformation values for the three copies, and given all the fields, specially the three tetrads at the same point, we can reconstruct the local $SU(2)$ transformation that gave rise either to three local LB1 transformations or independently to three local LB2 spatial transformations. Because the theorems proved, represent local isomorphisms between either three LB1 groups and $SU(2)$ or independently three LB2 groups and $SU(2)$. It is our purpose in this paper both regarding the no-go theorems, simplifications in the system of differential equations, gauge invariant algorithms in order to find diagonalized expressions for stress-energy tensors, group theoretical results, etc, to prove analogous results for $SU(3)$. Implicitly for $SU(3) \times SU(2) \times U(1)$. This paper will be organized as follows. In section \ref{QS} we will introduce a specific and new technique in order to deal with ``gauge vectors'' in $SU(3) \times SU(2) \times U(1)$. Taking advantage of Cartan results \cite{RGLG} on products of exponentials in order to generate groups we will prove new group theorems involving isomorphisms of local $SU(3)$ and local tensor products of LB1 and LB2.  In section \ref{TSU3} the construction of the new tetrads will be presented and two new theorems regarding group isomorphisms will be proved. In section \ref{appli} three new gauge invariants will be introduced and a new algorithm for diagonalizing the $SU(3)$ stress-energy tensor will be developed. There are many new mathematical and geometrical elements to discuss in this paper for the case $SU(3) \times SU(2) \times U(1)$, that are not present in the $SU(2) \times U(1)$ case, such that this work is not only worthwhile but also necessary.

\section{Quotient Space}
\label{QS}

In the case we are studying, that is local $SU(3)$, we cannot proceed as in the $U(1)$ or $SU(2)$ case. There are involved $3 \times 3$ matrices and we have to develop a different strategy. We will use the notion of quotient space, in particular the known relation \cite{MN}$^{,}$\cite{CI},

\begin{equation}
SU(3) / SU(2) \cong S^{5}\ . \label{quotiantu}
\end{equation}

But let us first understand what we will do through a simple example,

\begin{equation}
SO(3) / SO(2) \cong S^{2}\ . \label{quotianto}
\end{equation}

It is possible to generate all $SO(3)$ transformations by fixing a direction in $S^{2}$, that is choosing a unit vector in one special direction, and then performing all possible $SO(2)$ transformations in the orthogonal plane. If we repeat this process for all directions in $S^{2}$, then we would be spanning the whole $SO(3)$ group of local transformations. We can use this notion in order to implement a similar idea for the $SU(3)$ case which is not possible to visualize like the $SO(3)$ procedure.
Let us choose as our analog to $SO(2)$, the $SU(2)$ subalgebra to the $SU(3)$ algebra generated by the standard ``T'' elements \cite{GMS}. Let us call these three generators $X_{1}, X_{2}, X_{3}$. They obey the commutation relations $[X_{i},X_{j}] = \imath\:\epsilon_{ijk}\:X_{k}$, where the sum convention was applied on $k$. Let us call $\sigma^{\mu} = (\bf{1}, X_{1}, X_{2}, X_{3})$, where $\sigma^{i} = X_{i}$ are the $3 \times 3$ $SU(2)$ subalgebra generators of $SU(3)$,  essentially the Pauli matrices for $i = 1\cdots3$. Then it is clear that if we call $S_{ROT} = \exp \{ (\imath/2)\:\sum_{i=1}^{3} \theta_{i}\:X_{i} \}$, then

\begin{equation}
S_{ROT}\:\sigma^{\mu}\:S_{ROT}^{\dag} = \Lambda^{\mu}_{\:\:\:\nu}\:\sigma^{\nu}\ .\label{sigmatr}
\end{equation}

where $\theta_{i}$ are local scalars. Equation (\ref{sigmatr}) means that $\sigma^{\mu}$ transforms following the same pattern as vectors. In (\ref{sigmatr}), the matrices $S_{ROT}$ are local, as well as $\Lambda^{\mu}_{\:\:\:\nu}$. Since the group $SU(2)$ is homomorphic to $SO(3)$, they just represent local space rotations. For a more detailed discussion see the second appendix in \cite{A2}. Next, we build the local $SU(3)$ group element object $S = \exp \{ (\imath/2)\:\sum_{i=4}^{8} \theta_{i}\:X_{i} \}$. The $\theta_{i}, i=4\cdots8$ are all local scalars. The $X_{i}, i=4\cdots8$ are the remaining five $SU(3)$ group generators. The first three were associated to the $SU(2)$ subalgebra. Therefore,

\begin{equation}
S\:S_{ROT}\:S^{\dag}\:S\:\sigma^{\mu}\:S^{\dag}\:S\:S_{ROT}^{\dag}\:S^{\dag} = \Lambda^{\mu}_{\:\:\:\nu}\:S\:\sigma^{\nu}\:S^{\dag}\ .\label{sigmatrrot}
\end{equation}

We call $S$ the $SU(3)$ local group coset elements that represent a direction in $S^{5}$ about which $SU(2)$ ``rotations'' are performed, see section \ref{appendixI} for details. We remind ourselves that for every choice of a vector in $S^{5}$ we perform all possible local gauge $SU(2)$ transformations. We repeat this process for every vector in $S^{5}$. Now, every possible direction in $S^{5}$ is represented by a different local $SU(3)$ coset element $S$. They span a space, a manifold, not a group by themselves. In the end we translate every $SU(3)$ local gauge transformation into a product of transformations, following the ideas of Cartan, see chapter VII in\cite{RGLG}. In our notation every local $SU(3)$ group element will be written as $S\:S_{ROT}$. This way we will be able to ``link'' the local $SU(3)$ to the local $SU(2)$ and establish local group isomorphisms between the local $SU(3)$ and the local tensor products of LB1 or LB2. Keeping the above notation, if we call $\sigma_{5}=S\:\sigma^{\mu}\:S^{\dag}$, we can rewrite equation (\ref{sigmatrrot}),

\begin{equation}
(S\:S_{ROT}\:S^{\dag})\:\sigma_{5}^{\mu}\:(S\:S_{ROT}\:S^{\dag})^{\dag} = \Lambda^{\mu}_{\:\:\:\nu}\:\sigma_{5}^{\nu}\ .\label{sigmatrrot}
\end{equation}

It is also clear that $[\sigma_{5}^{i},\sigma_{5}^{j}] = \imath\:\epsilon_{ijk}\:\sigma_{5}^{k}$.\\

\section{Tetrads in $SU(3)$}
\label{TSU3}

It would be now appropriate to introduce the system of equations that is at the foundation of this work.

\begin{eqnarray}
R_{\mu\nu} &=& T^{(ymsu3)}_{\mu\nu} + T^{(ymsu2)}_{\mu\nu} + T^{(em)}_{\mu\nu}\label{eyme}\\
f^{\mu\nu}_{u(1)\:\:\:\:;\nu} &=& 0 \label{EM1}\\
\ast f^{\mu\nu}_{u(1)\:\:\:\:;\nu} &=& 0 \label{EM2}\\
f^{k\mu\nu}_{su(2)\:\:\:\:\:\:\:\mid \nu} &=& 0 \label{ymsu21}\\
\ast f^{k\mu\nu}_{su(2)\:\:\:\:\:\:\:\mid \nu} &=& 0 \label{ymsu22}\\
f^{p\mu\nu}_{su(3)\:\:\:\:\:\:\:\mid \nu} &=& 0 \label{ymsu31}\\
\ast f^{p\mu\nu}_{su(3)\:\:\:\:\:\:\:\mid \nu} &=& 0 \ . \label{ymsu32}
\end{eqnarray}

where the internal index $k$ is an $SU(2)$ index running from $k=1\cdots3$, while $p$ is an $SU(3)$ index running from $p=1\cdots8$.  The symbol ``;'' stands for the usual covariant derivative associated with the metric tensor $g_{\mu\nu}$, while $\mid$ stands for gauge covariant derivative. The tensors to the right of equation (\ref{eyme}) are the $SU(3)$, $SU(2)$ and $U(1)$ standard stress-energy tensors \cite{MC2}. First of all let us say that it is clear by now that we can proceed to build the tetrad skeletons following a similar procedure as in section ``Extremal field in $SU(2)$ geometrodynamics'' in paper \cite{A2}. It would be redundant to repeat it here. We will just introduce the new tetrad that we build following the constructions in papers \cite{A}$^{,}$\cite{A2}$^{,}$\cite{A3},

\begin{eqnarray}
Q_{(1)}^{\mu} &=& \Omega^{\mu\lambda}\:\Omega_{\rho\lambda}\:X^{\rho}
\label{S1}\\
Q_{(2)}^{\mu} &=& \sqrt{-Q_{ym}/2} \: \Omega^{\mu\lambda} \: X_{\lambda}
\label{S2}\\
Q_{(3)}^{\mu} &=& \sqrt{-Q_{ym}/2} \: \ast \Omega^{\mu\lambda} \: Y_{\lambda}
\label{S3}\\
Q_{(4)}^{\mu} &=& \ast \Omega^{\mu\lambda}\: \ast\Omega_{\rho\lambda}
\:Y^{\rho}\ ,\label{S4}
\end{eqnarray}

where $Q_{ym} = \Omega_{\mu\nu}\:\Omega^{\mu\nu}$ and $\Omega_{\mu\nu}\:\ast \Omega^{\mu\nu} = 0$. Then, let us define the ``gauge'' vector for our $SU(3)$ case,

\begin{equation}
X^{\sigma} = Y^{\sigma} = Tr[\Sigma^{\alpha\beta}\:S_{\alpha}^{\:\:\rho}\: S_{\beta}^{\:\:\lambda}\:\ast \epsilon_{\rho}^{\:\:\sigma}\:\ast \epsilon_{\lambda\tau}\:A^{\tau}] \ .\label{gaugev}
\end{equation}

$\Sigma^{\alpha\beta}$ is the antisymmetric object defined exactly as in \cite{A2} for the $X_{1}, X_{2}, X_{3}$ which are $3 \times 3$ matrices (see appendix II in reference \cite{A2}). $S_{\alpha}^{\:\:\rho}$ are the local $SU(2)$ tetrads defined exactly as in \cite{A2} for a $SU(2)$ gauge vector. Let us remember that $A^{\tau}$ in (\ref{gaugev}) is a $SU(3)$ gauge vector, a $3 \times 3$ matrix. The structure $S_{\alpha}^{\:\:[\rho}\: S_{\beta}^{\:\:\lambda]}\:\ast \epsilon_{\rho}^{\:\:\sigma}\:\ast \epsilon_{\lambda\tau}$ is invariant under $SU(2)$ local gauge transformations. Essentially, because of the $SU(2)$ extremal field property \cite{A}$^{,}$\cite{A2}$^{,}$\cite{MW}, $\epsilon_{\mu\sigma}\:\ast \epsilon^{\mu\tau} = 0$. Leaving thus in the contraction with $\ast \epsilon_{\rho\sigma}\:\ast \epsilon_{\lambda\tau}$, and because of property $\epsilon_{\mu\sigma}\:\ast \epsilon^{\mu\tau} = 0$, only the antisymmetric object $S_{2}^{\:\:[\rho}\:S_{3}^{\:\:\lambda]}$, which is locally $SU(2)$ gauge invariant. Let us remember that the object $\Sigma^{\alpha\beta}$ is antisymmetric and contracted with the $SU(2)$ tetrads as $\Sigma^{\alpha\beta}\:S_{\alpha}^{\:\:\rho}\: S_{\beta}^{\:\:\lambda}$ inside the local gauge vector (\ref{gaugev}). We already knew (see the second appendix in \cite{A2}) that the $SU(2)$ tetrads $S_{\beta}^{\:\:\lambda}$ were themselves invariant under local $U(1)$ gauge transformations \cite{A2}. We mean by this last remark, their tetrad skeletons and specially defined gauge vectors. Therefore, the $SU(3)$ gauge vectors $X^{\sigma} = Y^{\sigma}$ are locally invariant under $SU(2) \times U(1)$ local gauge transformations. This is fundamental since it enables us to introduce local $SU(3)$ tetrad gauge transformations independently of $SU(2) \times U(1)$ local gauge transformations and talk about $SU(3) \times SU(2) \times U(1)$ Yang-Mills geometrodynamics. It is important not to get confused by the product of exponentials in $SU(3)$ where one of the factors is a $SU(2)$ element of the subalgebra on one hand, and ``pure'' $SU(2)$ gauge transformations as we studied in paper \cite{A2}. The $SU(2)$ subalgebra ``operates'' through the gauge vector (\ref{gaugev}), while the ``pure'' $SU(2)$ gauge transformations get into play through a ``pure'' $SU(2)$ gauge vector as the one developed in paper \cite{A2}. When these two different gauge vectors for $SU(2)$ and $SU(3)$ tetrads are added as a possible choice of gauge vector in order to gauge the tetrad vectors, their inherent transformations are independent and the transformations LB1 and LB2 they induce commute between themselves, see Appendix III in reference \cite{A2}. Next we proceed to study the tetrad gauge transformations of the gauge vector (\ref{gaugev}). It is straightforward to notice that we can borrow all the analysis previously done in the section gauge geometry in \cite{A2}. Nonetheless it is important to pay attention to the nature of the gauge transformed local gauge vector (\ref{gaugev}) under the sequence of local $SU(3)$ transformations introduced in section \ref{QS}. This way, we are able to study the local gauge transformation of gauge vector (\ref{gaugev}) under any local $SU(3)$ gauge transformation since the product of exponentials covers all of the local $SU(3)$. Then, we proceed to transform (\ref{gaugev}) in a sequence, first with $S_{ROT}$, the local subalgebra group element, and then with $S$, the local coset representative,


\begin{eqnarray}
Tr[S\:S_{ROT}\:\Sigma^{\alpha\beta}\:S_{ROT}^{-1}\:S^{-1}\:S_{\alpha}^{\:\:\rho}\: S_{\beta}^{\:\:\lambda}\:\ast \epsilon_{\rho\sigma}\:\ast \epsilon_{\lambda\tau}\:A^{\tau}] + \nonumber \\ {\imath \over g} \:Tr[S\:S_{ROT}\:\Sigma^{\alpha\beta}\:S_{ROT}^{-1}\:S^{-1}\:S_{\alpha}^{\:\:\rho}\: S_{\beta}^{\:\:\lambda}\:\ast \epsilon_{\rho\sigma}\:\ast \epsilon_{\lambda\tau}\:\partial^{\tau}\:(S\:S_{ROT})\:(S\:S_{ROT})^{-1}]
\label{S1R}
\end{eqnarray}

For the sake of simplicity we are using the notation, $\Lambda^{(-1)\,\alpha}_{\:\:\:\:\:\:\:\:\:\:\:\:\delta} = \tilde{\Lambda}^{\alpha}_{\:\:\:\delta}$. Now, we can make use of the local transformation properties of the objects $\Sigma^{\alpha\beta}$, see section appendix II in reference \cite{A2}, and write,

\begin{eqnarray}
Tr[\tilde{\Lambda}^{\alpha}_{\:\:\:\delta}\:\tilde{\Lambda}^{\beta}_{\:\:\:\gamma}\:S\:\Sigma^{\delta\gamma}\:S^{-1}\:S_{\alpha}^{\:\:\rho}\: S_{\beta}^{\:\:\lambda}\:\ast \epsilon_{\rho\sigma}\:\ast \epsilon_{\lambda\tau}\:A^{\tau}] + \nonumber \\ {\imath \over g} \: Tr[\tilde{\Lambda}^{\alpha}_{\:\:\:\delta}\:\tilde{\Lambda}^{\beta}_{\:\:\:\gamma}\:S\:\Sigma^{\delta\gamma}\:S^{-1}\:S_{\alpha}^{\:\:\rho}\: S_{\beta}^{\:\:\lambda}\:\ast \epsilon_{\rho\sigma}\:\ast \epsilon_{\lambda\tau}\:\partial^{\tau}\:(S\:S_{ROT})\:(S\:S_{ROT})^{-1}] \ .
\label{S1L}
\end{eqnarray}

Making use now of the notation introduced in section \ref{QS} we can rewrite expression (\ref{S1L}) as,

\begin{eqnarray}
Tr[\tilde{\Lambda}^{\alpha}_{\:\:\:\delta}\:\tilde{\Lambda}^{\beta}_{\:\:\:\gamma}\:\Sigma_{5}^{\delta\gamma}\:S_{\alpha}^{\:\:\rho}\: S_{\beta}^{\:\:\lambda}\:\ast \epsilon_{\rho\sigma}\:\ast \epsilon_{\lambda\tau}\:A^{\tau}] + \nonumber \\ {\imath \over g} \: Tr[\tilde{\Lambda}^{\alpha}_{\:\:\:\delta}\:\tilde{\Lambda}^{\beta}_{\:\:\:\gamma}\:\Sigma_{5}^{\delta\gamma}\:S_{\alpha}^{\:\:\rho}\: S_{\beta}^{\:\:\lambda}\:\ast \epsilon_{\rho\sigma}\:\ast \epsilon_{\lambda\tau}\:\partial^{\tau}\:(S\:S_{ROT})\:(S\:S_{ROT})^{-1}] \ .
\label{S1L2}
\end{eqnarray}

The object $\Sigma_{5}^{\delta\gamma}$ is a $\Sigma^{\delta\gamma}$ object transformed by a local coset representative. The same element for all the local $SU(2)$ subalgebra. This locally gauge transformed gauge vector represented by equation (\ref{S1L2}) from a geometrical point of view has the following meaning \cite{A}$^{,}$\cite{A2}. Let us focus on blade one, on blade two the analysis is analogous.

Our first conclusion from the results above, is that $SU(3)$ local gauge transformations, generate the composition of two transformations. First, there is a local tetrad transformation, generated by a locally inertial coordinate transformation $\tilde{\Lambda}^{\alpha}_{\:\:\:\delta}$, of the $SU(2)$ tetrads $S_{\alpha}^{\rho}$ inside the gauge vector. Second, the normalized tetrad vectors that generate blade one, which the vector (\ref{gaugev}) is gauging, undergo a LB1 transformation on the blade they generate because of the second term in equation (\ref{S1L2}). The two normalized vectors, that generate blade one, end up on the same blade one generated by the original normalized generators of the blade, $\left({Q_{(1)}^{\mu} \over \sqrt{-Q_{(1)}^{\nu}\:Q_{(1)\nu}}}, {Q_{(2)}^{\mu} \over \sqrt{Q_{(2)}^{\nu}\:Q_{(2)\nu}}}\right)$, after the LB1 local transformation (see the tetrad vectors introduced in (\ref{S1}-\ref{S4})). This second transformation is generated by the second line in (\ref{S1L2}). Therefore, the gauge invariance of the metric tensor is assured as was discussed in papers \cite{A}$^{,}$\cite{A2}. We can continue making relevant remarks about these tetrad transformations. Within the set of LB1 tetrad transformations  of the pair $\left({Q_{(1)}^{\mu} \over \sqrt{-Q_{(1)}^{\nu}\:Q_{(1)\nu}}}, {Q_{(2)}^{\mu} \over \sqrt{Q_{(2)}^{\nu}\:Q_{(2)\nu}}}\right)$, there is an identity transformation that corresponds to the identity in $SU(3)$. To every LB1 tetrad transformation, which in turn is generated by $S\:S_{ROT}$ in $SU(3)$, there corresponds an inverse, generated by $S_{ROT}^{-1}\:S^{-1}$. We will prove in \ref{appendixII} that the inverse $S_{ROT}^{-1}\:S^{-1}$ can be reexpressed as a coset element times a subalgebra element, that is, can be rewritten in the same original coset parametrization. We observe also the following. Since locally inertial coordinate Lorentz transformations $\tilde{\Lambda}^{\alpha}_{\:\:\:\delta}$ of the $SU(2)$ tetrads $S_{\alpha}^{\rho}$ in general do not commute, then the locally $SU(3)$ tetrad generated transformations are non-Abelian. The non-Abelianity of $SU(3)$ is mirrored by the non-commutativity of these locally inertial coordinate transformations $\tilde{\Lambda}^{\alpha}_{\:\:\:\delta}$, which are esentially local space rotations. The key role in this non-commutativity is played by the object $\Sigma^{\alpha\beta}$, that translates the local $SU(2)$ subalgebra factor in local $SU(3)$ gauge transformations, into locally inertial Lorentz transformations. Another issue of relevance is related to the analysis of the ``memory'' of these transformations. As we did in paper \cite{A2} we would like to know explicitly, if a second LB1 tetrad transformation, generated by a local gauge transformation $S_{2}\:S_{ROT2}$, is going to ``remember'' the existence of the first one, generated by $S_{1}\:S_{ROT1}$. To this end and following the lines in \cite{A2}, let us just write for instance the vector $Q_{(1)}^{\:\mu}$ after these two gauge transformations,


\begin{eqnarray}
\lefteqn{Q_{(1)}^{\:\mu} \rightarrow  \Omega^{\mu\nu}\:\Omega^{\sigma}_{\:\:\nu}\: Tr[\tilde{\Lambda}^{\alpha}_{2\:\:\kappa}\:\tilde{\Lambda}^{\beta}_{2\:\:\Omega}\:\tilde{\Lambda}^{\kappa}_{1\:\:\delta}\:
\tilde{\Lambda}^{\Omega}_{1\:\:\gamma}\:(S_{5}\:\Sigma^{\delta\gamma}\:S_{5}^{-1})\:S_{\alpha}^{\:\:\rho}\: S_{\beta}^{\:\:\lambda}\:\ast \epsilon_{\rho\sigma}\:\ast \epsilon_{\lambda\tau}\:A^{\tau}] +} \nonumber \\
&&{\imath \over g} \:\Omega^{\mu\nu}\:\Omega^{\sigma}_{\:\:\nu}\: Tr[\tilde{\Lambda}^{\alpha}_{2\:\:\kappa}\:\tilde{\Lambda}^{\beta}_{2\:\:\Omega}\:\tilde{\Lambda}^{\kappa}_{1\:\:\delta}\:
\tilde{\Lambda}^{\Omega}_{1\:\:\gamma}\:(S_{5}\Sigma^{\delta\gamma}\:S_{5}^{-1})\:S_{\alpha}^{\:\:\rho}\: S_{\beta}^{\:\:\lambda}\:\ast \epsilon_{\rho\sigma}\:\ast \epsilon_{\lambda\tau}\:\partial^{\tau}\:(S_{1}\:S_{ROT1})\:(S_{1}\:S_{ROT1})^{-1}] + \nonumber \\
&&{\imath \over g} \:\Omega^{\mu\nu}\:\Omega^{\sigma}_{\:\:\nu}\: Tr[\tilde{\Lambda}^{\alpha}_{2\:\:\delta}\:\tilde{\Lambda}^{\beta}_{2\:\:\gamma}\:(S_{2}\:\Sigma^{\delta\gamma}\:S_{2}^{-1})\:S_{\alpha}^{\:\:\rho}\:
S_{\beta}^{\:\:\lambda}\:\ast \epsilon_{\rho\sigma}\:\ast \epsilon_{\lambda\tau}\:\partial^{\tau}\:(S_{2}\:S_{ROT2})\:(S_{2}\:S_{ROT2})^{-1}] \ .
\label{S1DOUBLETR}
\end{eqnarray}

where $S_{5} = S_{1}\:S_{ROT1}\:S_{2}\:S_{ROT1}^{-1}$. It is clear that in the third line in (\ref{S1DOUBLETR}) neither $S_{1}$ or $S_{ROT1}$ are present. This third line represents the second LB1 transformation. In the second line through $S_{5}$, only $S_{2}$ is present with regards to the second $S_{2}\:S_{ROT2}$ local $SU(3)$ gauge transformation. Now, the point is that $S_{2}$ is the coset representative for the second $SU(3)$ local gauge transformation. It is given by a vector in $S^{5}$, the quotient space, see section \ref{QS}.



Therefore, the second subgroup spanned by the elements $S_{ROT2}$, the local subgroup to $SU(3)$, is not present in the second line that represents the first LB1 local transformation. $S_{2}$ is one coset element representing a fixed direction in the local $S^{5}$, just one for the whole of the local $SU(2)$ subgroup spanned by all the $S_{ROT2}$. Therefore the second local $SU(3)$ gauge transformation is only present through the local equivalence class coset representative. With regard to the object $S_{ROT1}\:S_{2}\:S_{ROT1}^{-1}$, we will study its properties in \ref{appendixII}. We will prove that this object is a local coset representative by itself. Therefore $S_{5}$ is nothing but the product of two coset representatives.

We can notice that the second term contains the same $SU(2)$ transformed tetrads as the first one. Therefore, when we compare these two terms, it is straightforward to see that it is not possible, after the second gauge transformation, from these transformed $SU(2)$ tetrads, to ``remember'' any relative change associated to the second gauge transformation.

In addition, the second line in (\ref{S1DOUBLETR}) contains in the derivative only $S_{1}\:S_{ROT1}$, and the third line contains only $S_{2}\:S_{ROT2}$. This means that the second LB1 tetrad transformation on blade one is not going to remember the first one. The algebra underlying these statements can be followed through \cite{A}. Again, as reasoned in \cite{A2}, another way of thinking of (\ref{S1DOUBLETR}) is by first performing two successive local Lorentz transformations of the $SU(2)$ tetrad in the first line, and second, by performing two successive $LB1$ tetrad transformations in the second and third line.

We could repeat exactly the remainder of the analysis done in paper \cite{A2} in section ``gauge geometry'' and reach similar conclusions, the only specific and distinctive issue of surjectivity in the local $SU(3)$ case that should be highlighted and discussed carefully and that we will study in detail in section \ref{appendixIII} . Finally we are able to state two new theorems.

\newtheorem {guesslb1} {Theorem}
\newtheorem {guesslb2}[guesslb1] {Theorem}
\begin{guesslb1}
The mapping between the local gauge group of transformations $SU(3)$ and the tensor product of the eight local groups of LB1 transformations is isomorphic. \end{guesslb1}

Following analogously the reasoning laid out in \cite{A}, in addition to the ideas above, we can also state,

\begin{guesslb2}
The mapping between the local gauge group of transformations $SU(3)$ and the tensor product of the eight local groups of LB2 transformations is isomorphic. \end{guesslb2}

The group $SU(3)$ is connected to the identity making unnecessary to discuss isomorphisms of group sheet components as in the $SU(2)$ case. It is very important that we make a clarification on an issue that might give rise to confusion.
In theorems (1-2) we are considering eight copies of the same spacetime, and a different tetrad at the same point in each spacetime copy. These eight tetrads have a similar extremal field-gauge vector structure as in (\ref{S1}-\ref{S4}). They are normalized and a choice of gauge vector has been made. But they are not the same. They could be Lorentz transformed into each other, under non-trivial Lorentz spatial rotations, for instance. We know from manuscript \cite{A2} that a local Lorentz transformation of a tetrad with an extremal field-gauge vector structure transforms into another tetrad with a similar extremal field-gauge vector structure. Therefore, we are considering eight local tetrads at the same spacetime point which are not the same for different copies. The local planes one and two will be tilted with respect to each other. This is what we mean by eight LB1 or LB2 groups under tensor product. Now, from a practical point of view what we also mean by these theorems is that we are able to reconstruct the original local $SU(3)$ gauge transformation by knowing for example, the boosts in the LB1 case, or the spatial rotations in the LB2 case, for the tetrad local transformations at the point under consideration. By knowing the local Lorentz transformation values for the eight copies, and given all the fields, specially the eight tetrads at the same point, we can reconstruct the local $SU(3)$ transformation that gave rise either to eight local LB1 transformations or independently to eight local LB2 spatial transformations. Because the theorems proved, represent local isomorphisms between either eight LB1 groups and $SU(3)$ or independently eight LB2 groups and $SU(3)$.

\section{Applications}
\label{appli}

\subsection{Gauge invariants}
\label{Gauge invariants}

The analysis of gauge invariants and stress-energy tensor diagonalization will proceed in an analogous way as in paper \cite{A2}, section Applications. The invariants will be presented in a similar fashion as in paper \cite{A2}, however, we would like to deepen in the analysis of gauge invariant diagonalization on certain details that deserve particular attention for the $SU(3)$ case. As in the local $SU(2)$ case we would like to introduce new gauge invariant objects built out of the components of the stress-energy tensor. Only in this section when we write $T_{\mu\nu}$ we mean $T^{(ymsu3)}_{\mu\nu}$, once again we just do not want to overload the equations with notation. Given the tetrad $W_{(o)}^{\mu}$, $W_{(1)}^{\mu}$, $W_{(2)}^{\mu}$, $W_{(3)}^{\mu}$, (no confusion should arise with vector $E_{(3)}^{\:\:\rho} = W^{\rho}$ which is just one vector in the electromagnetic tetrad) which we consider to be the normalized version of $Q_{(1)}^{\mu}$, $Q_{(2)}^{\mu}$, $Q_{(3)}^{\mu}$, $Q_{(4)}^{\mu}$, (\ref{S1}-\ref{S4}) ,we perform the local $SU(3)$ gauge transformations on blades one and two,

\begin{eqnarray}
\tilde{W}_{(o)}^{\mu} &=& \cosh\phi\:W_{(o)}^{\mu} + \sinh\phi\:W_{(1)}^{\mu}\label{GT1}\\
\tilde{W}_{(1)}^{\mu} &=& \sinh\phi\:W_{(o)}^{\mu} + \cosh\phi\:W_{(1)}^{\mu}\label{GT2}\\
\tilde{W}_{(2)}^{\mu} &=& \cos\psi\:W_{(2)}^{\mu} - \sin\psi\:W_{(3)}^{\mu}\label{GT3}\\
\tilde{W}_{(3)}^{\mu} &=& \sin\psi\:W_{(2)}^{\mu} + \cos\psi\:W_{(3)}^{\mu}\ .\label{GT4}
\end{eqnarray}

The scalars $\phi$ and $\psi$ are local. It is a matter of algebra to prove that the following objects are invariant under the set of transformations (\ref{GT1}-\ref{GT4}),

\begin{eqnarray}
\lefteqn{ \left(\:W_{(0)}^{\mu}\:T_{\mu\nu}\:W_{(0)}^{\nu}\right)\:W_{(0)}^{\lambda}\:W_{(0)}^{\rho} - \left(\:W_{(0)}^{\mu}\:T_{\mu\nu}\:W_{(1)}^{\nu}\right)\:\left[W_{(0)}^{\lambda}\:W_{(1)}^{\rho} + W_{(0)}^{\rho}\:W_{(1)}^{\lambda}\right] + } \nonumber \\
&&\left(\:W_{(1)}^{\mu}\:T_{\mu\nu}\:W_{(1)}^{\nu}\right)\:W_{(1)}^{\lambda}\:W_{(1)}^{\rho}  \label{GI1} \\
&&-\left(\:W_{(0)}^{\mu}\:T_{\mu\nu}\:W_{(2)}^{\nu}\right)\:\left[W_{(0)}^{\lambda}\:W_{(2)}^{\rho} + W_{(0)}^{\rho}\:W_{(2)}^{\lambda}\right]
-\left(\:W_{(0)}^{\mu}\:T_{\mu\nu}\:W_{(3)}^{\nu}\right)\:\left[W_{(0)}^{\lambda}\:W_{(3)}^{\rho} + W_{(0)}^{\rho}\:W_{(3)}^{\lambda}\right] + \nonumber \\
&&\left(\:W_{(1)}^{\mu}\:T_{\mu\nu}\:W_{(2)}^{\nu}\right)\:\left[W_{(1)}^{\lambda}\:W_{(2)}^{\rho} + W_{(1)}^{\rho}\:W_{(2)}^{\lambda}\right] + \
\left(\:W_{(1)}^{\mu}\:T_{\mu\nu}\:W_{(3)}^{\nu}\right)\:\left[W_{(1)}^{\lambda}\:W_{(3)}^{\rho} + W_{(1)}^{\rho}\:W_{(3)}^{\lambda}\right] \label{GI2}\\
&&\left(\:W_{(2)}^{\mu}\:T_{\mu\nu}\:W_{(2)}^{\nu}\right)\:W_{(2)}^{\lambda}\:W_{(2)}^{\rho} + \left(\:W_{(2)}^{\mu}\:T_{\mu\nu}\:W_{(3)}^{\nu}\right)\:\left[W_{(2)}^{\lambda}\:W_{(3)}^{\rho} + W_{(2)}^{\rho}\:W_{(3)}^{\lambda}\right] + \nonumber \\
&&\left(\:W_{(3)}^{\mu}\:T_{\mu\nu}\:W_{(3)}^{\nu}\right)\:W_{(3)}^{\lambda}\:W_{(3)}^{\rho}\ .\label{GI3}
\end{eqnarray}

The ensuing discussion, now for the local $SU(3)$ case, about the gauge transformation properties of the objects (\ref{GI1}-\ref{GI3}) is similar to the one presented in paper \cite{A2}. Using normalized tetrads, and under tetrad transformations of the kind (\ref{GT1}-\ref{GT4}), the objects (\ref{GI1}-\ref{GI3}) will remain invariant. The point is that the transformations (\ref{GT1}-\ref{GT4}), represent local $SU(3)$ gauge transformations of the tetrad vectors, or tetrad gauge generated trasformations \cite{A}$^{,}$\cite{A2}. It is the way in which the normalized version of tetrad vectors (\ref{S1}-\ref{S4}) transform on blades one and two under locally generated $SU(3)$ gauge transformations. The tensor $T_{\mu\nu}$ is gauge invariant by itself as we already know. Then these are true new local $SU(3)$ gauge invariants under (\ref{GT1}-\ref{GT4}). Again as in paper \cite{A2} we remember what happens with the objects (\ref{GI1}-\ref{GI3}) when we perform discrete gauge transformations on blade one. The objects remain invariant under a tetrad full inversion on blade one. However, under the discrete transformation represented by equations (64-65) in \cite{A}, while objects (\ref{GI1}) and (\ref{GI3}) remain invariant, object (\ref{GI2}) changes in a global sign (gets multiplied globally by $-1$). Therefore, and now for the local $SU(3)$ case we can say that objects (\ref{GI1}) and (\ref{GI3}) are true and new gauge invariants, while object (\ref{GI2}) is invariant under boosts generated gauge transformations on blade one, rotations on blade two, full inversions on blade one, but gets multiplied by $-1$ under the discrete gauge generated transformation (64-65) \cite{A} on blade one. Once again we will make use of these gauge properties of objects (\ref{GI1}-\ref{GI3}) in the next section that deals with diagonalization of the stress-energy tensor, this time in the local $SU(3)$ case.

\subsection{Diagonalization of the stress-energy tensor}

We proceed now to extend to the non-Abelian $SU(3)$ case the algorithm for the diagonalization of the stress-energy tensor. It is worth mentioning once again that we will present one method, but there are others, all equivalent of course. Another method that we call multiple extremal representation of the gravitational field will be introduced in an upcoming paper. In the previous section \ref{Gauge invariants} we found that we can build with the stress-energy tensor and the new tetrads, three objects that are locally gauge invariant. This is a mathematical truth that can be easily checked. Then, as in paper \cite{A2}, we might ask about the usefulness of the existence of these three new gauge invariant objects, and our answer is the following. These three new local gauge invariant objects allow us to connect gauge invariance with three different blocks in the stress-energy tensor. One block off-diagonal and two diagonal blocks, separately. Therefore these three new gauge invariant objects will guide us in establishing a local gauge invariant process of diagonalization of the stress-energy tensor. Their existence means that we can block diagonalize the stress-energy tensor in a gauge invariant way, locally. As in paper \cite{A2} we start by putting forward a generalized duality transformation for non-Abelian fields, this time for the local $SU(3)$ situation. For instance we might choose,

\begin{equation}
\varepsilon_{\mu\nu} =  Tr[\vec{n}\: \cdot \: f_{\mu\nu} - \vec{l}\: \cdot \: \ast f_{\mu\nu}] \ ,\label{gendual}
\end{equation}

where $f_{\mu\nu} = f^{a}_{\mu\nu}\:X^{a}$, $\:\:\vec{n} = n^{a}\:X^{a}$ and $\vec{l} = l^{a}\:X^{a}$ are vectors in the eight-dimensional internal space. The $\cdot$ means product in internal space. $X^{a}$ are the $SU(3)$ generator matrices see section \ref{QS} and reference \cite{GMS} and the summation convention is applied on the internal index $a$. The vector components are defined as,

\begin{eqnarray}
\lefteqn{ \vec{n} = (\cos\theta_{1},\cos\theta_{2},\cos\theta_{3},0 \cdots 0) } \label{ISO1} \\
&&\vec{l} = (\cos\beta_{1},\cos\beta_{2},\cos\beta_{3},0 \cdots 0 ) \ , \label{ISO2}
\end{eqnarray}

where the vectors have eight components and all the six angles are local scalars that satisfy,

\begin{eqnarray}
\lefteqn{ \Sigma_{a=1}^{3} \cos^{2}\theta_{a} = 1 } \label{ISOSUM1} \\
&&\Sigma_{a=1}^{3} \cos^{2}\beta_{a} = 1 \ . \label{ISOSUM2}
\end{eqnarray}

We can notice from (\ref{ISO1}-\ref{ISO2}) that the only angles chosen in the eight-dimensional internal space to be different from zero are the ones associated to the internal local $SU(2)$ subalgebra. They are enough to carry out the diagonalization algorithm. In eight-dimensional internal space $\vec{n} = n^{a}\:X^{a}$ transforms under a local $SU(2)$ gauge transformation $S$, that belongs to the subgroup of $SU(3)$ as $S^{-1}\:\vec{n}\:S$, see chapter III in \cite{CMDWYCB} and also\cite{GMS}, and similar for $\vec{l} = l^{a}\:X^{a}$. The tensor $f_{\mu\nu} = f^{a}_{\mu\nu}\:X^{a}$ transforms as
$f_{\mu\nu} \rightarrow S^{-1}\:f_{\mu\nu}\:S$. Therefore $\varepsilon_{\mu\nu}$ is manifestly gauge invariant under any local $SU(2)$ gauge transformation that belongs to the subgroup of $SU(3)$. We can see from (\ref{ISO1}-\ref{ISO2}) and (\ref{ISOSUM1}-\ref{ISOSUM2}) that only four of the six angles in internal space are independent. Next, using similar notation to paper \cite{A2} we perform one more duality transformation,

\begin{equation}
\Omega_{\mu\nu} = \cos\alpha_{d} \:\: \varepsilon_{\mu\nu} -
\sin\alpha_{d} \:\: \ast \varepsilon_{\mu\nu} \ ,\label{diagdual}
\end{equation}

such that the extremal field (\ref{diagdual}) satisfies $\Omega_{\mu\rho}\:\ast\Omega^{\mu\nu}=0$, and the complexion $\alpha_{d}$ is defined by,

\begin{eqnarray}
\tan(2\alpha_{d}) = - \varepsilon_{\mu\nu}\:\ast \varepsilon^{\mu\nu} / \varepsilon_{\lambda\rho}\:\varepsilon^{\lambda\rho}\ .\label{compdd}
\end{eqnarray}

All the conclusions derived in \cite{A} are valid in this context and therefore exactly as in reference \cite{A}. Using the local antisymmetric tensor $\Omega_{\mu\nu}$, we can produce tetrad skeletons and with new gauge vectors $X_{d}^{\sigma}$ and $Y_{d}^{\sigma}$ we can build a new normalized tetrad. This new tetrad that we call $T_{\alpha}^{\mu}$ has four independent isoangles included in its definition, in the skeletons. There is also the freedom to introduce an LB1 and an LB2 local $SU(3)$ generated transformations on both blades by new angles $\phi_{d}$ and $\psi_{d}$ (through the gauge vectors $X_{d}^{\sigma}$ and $Y_{d}^{\sigma}$) which are not yet fixed and represent two more independent angles. They are truly going to be LB1 and LB2 local $SU(2)$ generated transformations on both blades, as we will see in a moment. Having six independent and undefined angles, we will use this freedom to choose them when fixing the six diagonalization conditions for the stress-energy tensor. It must be highlighted and stressed that since the local antisymmetric tensor $\Omega_{\mu\nu}$ is gauge invariant, then the tetrad vectors skeletons are locally $SU(2)$ gauge invariant. Let us notice that they are not local $SU(3)$ gauge invariants since we cannot produce in the eight-dimensional internal space a transformation law of general vectors that proceeds as $S^{-1}\:\vec{n}\:S$ for a general local $S$ that belongs to $SU(3)$. However, we can produce this local transformation law for isovectors in the three-dimensional subspace corresponding to the local $SU(2)$ subalgebra. We remind ourselves that this was a fundamental condition that we made in previous sections in order to ensure the metric invariance when performing LB1 and LB2 transformations. Then, we proceed to impose the diagonalization conditions,

\begin{eqnarray}
\lefteqn{ T_{o1} = T_{o}^{\mu}\:T_{\mu\nu}\: T_{1}^{\nu} = 0 } \label{diagcond1} \\
&&T_{o2} = T_{o}^{\mu}\:T_{\mu\nu}\: T_{2}^{\nu} = 0 \label{diagcond2} \\
&&T_{o3} = T_{o}^{\mu}\:T_{\mu\nu}\: T_{3}^{\nu} = 0 \label{diagcond3} \\
&&T_{12} = T_{1}^{\mu}\:T_{\mu\nu}\: T_{2}^{\nu} = 0 \label{diagcond4} \\
&&T_{13} = T_{1}^{\mu}\:T_{\mu\nu}\: T_{3}^{\nu} = 0 \label{diagcond5} \\
&&T_{23} = T_{2}^{\mu}\:T_{\mu\nu}\: T_{3}^{\nu} = 0 \ .\label{diagcond6}
\end{eqnarray}

Notice the following property of equations (\ref{diagcond2}-\ref{diagcond5}). These equations are invariant under any local $SU(3)$ transformations (\ref{GT1}-\ref{GT4}) and even more so under any local $SU(2)$ that belongs to the subgroup of $SU(3)$. This property ensures that block diagonalization will be locally gauge invariant under any local transformation of the kind (\ref{GT1}-\ref{GT4}) in the local subgroup. These are finally the six equations that locally define the six angles $\theta_{1},\:\theta_{2},\:\beta_{1},\:\beta_{2},\:\phi_{d},\:\psi_{d}$, for instance. The other two $\theta_{3},\:\beta_{3}$ are determined by equations (\ref{ISOSUM1}-\ref{ISOSUM2}) once the other six have already been determined through equations (\ref{diagcond1}-\ref{diagcond6}). Once the stress-energy tensor has been diagonalized, as we did in paper \cite{A2} we can study the gauge invariants (\ref{GI1}-\ref{GI3}). Always assuming that the local diagonalization process is possible, in the new gauge, the ``diagonal gauge'', determined by the new gauge angles already found, we observe that the object (\ref{GI2}) is going to be zero or null when written in terms of the new ``diagonal tetrad'' $T_{\alpha}^{\mu}$. Let us remember that this particular object was invariant under all gauge transformations except those which changed it by a global sign. Therefore, we conclude, if its components are all null in one gauge, in this case the ``diagonal gauge'', they all will be null in any other gauge. The other two objects will be maximally simplified since the off-diagonal terms in both of them will vanish in the ``diagonal gauge''. It is evident that the ``diagonal gauge'' might be a source of simplification in dealing with the field equations, and of course the inherent simplification in the geometrical analysis of any problem involving these kind of fields (\ref{eyme}-\ref{ymsu32}). We would like before the end to this section to highlight a possible source of confusion in our diagonalization algorithm. The diagonalization algorithm in paper \cite{A2} was such that the tetrad skeletons chosen in a similar fashion as in equation (\ref{gendual}) were local gauge invariants of the theory. The local gauge transformations belong to the $SU(2)$ group in that case. Then, we proceeded to write the six diagonalization equations similar to (\ref{diagcond1}-\ref{diagcond6}). In that case the analogous equations to (\ref{diagcond2}-\ref{diagcond5}) represented a gauge invariant block diagonal system. This is simply because under analogous transformations to (\ref{GT1}-\ref{GT4}) of the vectors $T_{\alpha}^{\mu}$, equations (\ref{diagcond2}-\ref{diagcond5}) just mix among themselves. Finally, the two LB1 and LB2 gauge transformations determined the diagonal gauge. In our present paper we cannot implement a completely analogous procedure. The reason is simply that we do not have local transformations in the eight-dimensional internal space that relate local rotations of unit vectors in eight-dimensional space with a transformation of the kind $S^{-1}\:\vec{n}\:S$, such that $S$ is any local $SU(3)$ gauge transformation. If this kind of relationship would exist, then the block diagonalization would truly be locally $SU(3)$ gauge invariant. But this is not the case. We can make the block diagonal procedure only $SU(2)$ gauge invariant. The freedom to introduce an LB1 and an LB2 local $SU(3)$ generated transformations on both blades by new angles $\phi_{d}$ and $\psi_{d}$ will also be limited to the local $SU(2)$ subgroup, otherwise we would alter the skeletons. By no means all of this means that we cannot build local $SU(3)$ gauge invariant skeletons. We can in many ways, for an analogous discussion see section III in paper \cite{A2}. The diagonalization process will then be only local $SU(2)$ invariant. Not $SU(3)$ gauge invariant.

\section{Conclusions}
\label{conclusions}

It is important to highlight the many efforts carried out during the past century in the context of grand field unification, see for example the paper \cite{CBDEL} and the full set of references therein. In the present work there are so many new mathematical and geometrical elements to discuss for the case $SU(3) \times SU(2) \times U(1)$, that are not present in the $SU(2) \times U(1)$ case, that make this work not only worthwhile but also necessary. Similar to the $U(1)$ case, the $SU(3)$ local gauge group of transformations associated with Yang-Mills fields, finds its counterpart in geometrical structures. To find this relation between gauge, and geometrical structures we study a key property that they have. This property was already analyzed for $U(1)$ and is related to the fact that in the Abelian environment associated with electromagnetic fields, the local gauge transformation of the tetrad vectors induces LB1 Lorentz transformation on blade one, such that the two vectors that generate this blade, remain on the blade after the transformation. Similar for rotations on blade two. This property is essential as far as we ask for the metric tensor to remain invariant under $U(1)$ transformations in the Abelian case. We demanded a similar property for the metric tensor in spacetimes where $SU(2)$ Yang-Mills fields are present and do the same for the $SU(3)$ case. That is the reason why we take on the task of finding tetrads that have transformation properties analogous to the Abelian, in this non-Abelian environment. Once we build these new tetrads in section \ref{TSU3}, we study their transformation properties. They have an inherent freedom in the choice of two vector fields. These two vector fields are available freedom in the construction of our tetrads. As in papers \cite{A}$^{,}$\cite{A2}, they are ``gauge'' by themselves, and they include ``gauge'' in their construction. It is this freedom the tool for translating the abstract internal local group of transformations into spacetime local groups of transformations. These vectors chosen for this particular example in $SU(3) \times SU(2) \times U(1)$ do this job.

The physical and geometrical significance of this work reside in the following issues.

\begin{enumerate}
\item For many decades there were many attempts at trying to establish structures that included the spacetime groups of transformations, and the internal groups of transformations, see the no-go papers \cite{SWNG}$^{,}$\cite{LORNG}$^{,}$\cite{CMNG} and the references in them. It was not known that the local internal groups of transformations were isomorphic to local spacetime groups of transformations. This situation led to believe that the ``internal'' were detached from the ``spacetime''. It was thought that local groups of gauge ``internal'' transformations do not act on spacetime objects. It was concluded then, that the generators of the internal and spacetime groups commute. In our works\cite{A}$^{,}$\cite{A2} and the present manuscript, we proved in an explicit and manifest fashion that the local ``internal'' groups $SU(3)$, $SU(2)$ and $U(1)$ are isomorphic to local ``spacetime'' groups of tetrad transformations on the local orthogonal planes of diagonalization of stress-energy tensors, that is LB1, LB2 or tensor products of them. Therefore the aforementioned assumptions and conclusions are not true simply because on one hand local Lorentz transformations do not commute in general. If the local groups of gauge transformations are isomorphic to local groups of tetrad transformations on special planes, then these local Lorentz transformations do not commute with Lorentz transformations on local planes tilted with respect to the local unique planes of stress-energy diagonalization. On the other hand because LB1 groups (local boosts plus discrete transformations) are isomorphic to LB2 groups (local spatial rotations) via the compact $SU(3) \times SU(2) \times U(1)$.

\item Let us briefly remind ourselves what we mean by tensor products of one-dimensional groups LB1 or LB2. We are able to reconstruct the original local $SU(3)$ gauge transformation by knowing for example, the boosts in the LB1 case, or the spatial rotations in the LB2 case, for the tetrad local transformations at the point under consideration, for eight copies of the same spacetime. We also consider a different tetrad at the same point in each spacetime copy. These eight tetrads have a similar extremal field-gauge vector structure as in (\ref{S1}-\ref{S4}). They are normalized and a choice of gauge vector has been made. But they are not the same. They could be Lorentz transformed into each other, under non-trivial Lorentz spatial rotations, for instance. We know from manuscript \cite{A2} that a local Lorentz transformation of a tetrad with an extremal field-gauge vector structure transforms into another tetrad with a similar extremal field-gauge vector structure. Therefore, we are considering eight local tetrads at the same spacetime point which are not the same for different copies. The local planes one and two will be tilted with respect to each other. This is what we mean by eight LB1 or LB2 groups under tensor product. By knowing the local Lorentz transformation values for the eight copies, and given all the fields, specially the eight tetrads at the same point, we can reconstruct the local $SU(3)$ transformation that gave rise either to eight local LB1 transformations or independently to eight local LB2 spatial transformations. Because the theorems proved, represent local isomorphisms between either eight LB1 groups and $SU(3)$ or independently eight LB2 groups and $SU(3)$.

\item We are settling the issue about the relation between the groups so far regarded as generating local ``internal'' transformations, and local ``spacetime'' transformations. The standard model has been designed on the pillar of gauge invariance. Finding this relation amounts to finding the relationship to the gravitational field. The relevant point is that we know now that the ``link'' between internal structures and spacetime structures is bridged by tetrads, and local gauge transformations are isomorphic to local tetrad transformations that explicitly leave the metric tensor invariant. This knowledge draws other results. New theorems in group theory. Once again one may ask \cite{DESISH} if particle multiplets can be associated to gravitational fields which are explicitly invariant under these groups of local transformations \cite{GMS}. The microparticles would then be tetrad gauge ``states'' of the gravitational fields related to each other through local ``rotations'' LB1, LB2, which in turn are generated by internal local gauge transformations. Therefore we are introducing for the first time and explicit ``link'' between the ``internal'' and the ``spacetime'', so far detached from each other. Then again we conjecture the possibility of microparticles to be associated to spacetimes since all the standard model symmetries can be realized in four-dimensional Lorentzian curved spacetimes.

\item In the first paper \cite{A} we proved that the group $U(1)$ is isomorphic to the local group of boosts plus discrete transformations on blade one that we called LB1. One of the discrete transformations is not Lorentzian \cite{A}. As the same group $U(1)$ is isomorphic to $SO(2)$, that we also called LB2 since it is related to local Lorentz tetrad rotations on blade two, then the group $SO(2)$ is isomorphic to the proper group on blade one plus discrete transformations. This is a fundamental result in group theory and in physics. We are simultaneously proving, and this is the point that we are emphasizing in this item, that there is an isomorphism between inertial states and gauge states of the gravitational fields locally. In our present paper we proved two new theorems. First, the local group of $SU(3)$ gauge transformations is isomorphic to the tensor product of eight LB1 groups. Second, the local group of $SU(3)$ gauge transformations is isomorphic to the tensor product of eight LB2 or $SO(2)$ groups. Then, the local compact $SU(3)$ is isomorphic to the tensor product of eight local non-compact LB1 groups (boosts plus discrete transformations). This is another fundamental result in group theory, in physics as well. As in the Abelian case discussed in \cite{A}, and also the non-Abelian case discussed in paper \cite{A2}, and again this is the point that we are emphasizing in this item, we proved again in this non-Abelian case that there is an isomorphism between inertial states and gauge states of the gravitational fields locally.

\end{enumerate}

In previous works we proved theorems about the relationship of local groups of gauge transformations and local groups of tetrad transformations in curved spacetimes. In particular $U(1)$ in reference \cite{A}, and $SU(2) \times U(1)$ in reference \cite{A2,A3}. In this work we proved similar results but now with a local gauge group of transformations $SU(3) \times SU(2) \times U(1)$. We found geometrical counterparts to gauge structures. The special kind of tetrads we found in paper \cite{A} provided a fundamental tool. However, duality transformations were elementary in the electromagnetic case. We had to find new more general local duality transformations \cite{DEHETEI}$^{,}$\cite{HETEI}$^{,}$\cite{DEGOHETEI} when developing an algorithm for diagonalization of the stress-energy tensor both in the $SU(2)$, and the $SU(3)$ cases. Duality transformations turned out to be a fundamental tool. The new gauge invariant objects (\ref{GI1}-\ref{GI3}) play a role in the diagonalization algorithm since by using them as a parallel guide we were able to establish a gauge invariant block diagonalization procedure. Once the skeletons are oriented at every point in spacetime, then we proceed to transform the tetrads through LB1 and LB2 ``rotations'', fixing the ``diagonal gauge''. In other words diagonalizing the two remaining blocks in the diagonal. In this way we devise the diagonalization algorithm as a geometrical process involving new duality transformations that relates tetrads, skeletons and gauge vectors with the stress-energy tensor diagonal final structure. It is relevant to underscore the many possible applications of the new tetrad techniques, for example in the field of Cosmology, see the paper \cite{CDEL} and the set of references in this work. We quote from \cite{RJ} ``However, if we put aside the issue of gravitational dynamics and focus only on the gravitational field variables, we can find many (notational) analogies to gauge fields. These analogies are useful for motivating and constructing gravitational counterparts to gauge theoretic entities. Correspondingly, aspects of general relativity can inform topics in gauge theory''. By establishing a link between the local gauge groups of transformations and local geometrical groups of transformations, like in papers \cite{A}$^{,}$\cite{A2,A3}, we are trying to bring the gauge theories into a geometric formulation. This is a paper about grand Standard Model gauge theories - General Relativity gravity unification and grand group unification in four-dimensional curved Lorentzian spacetimes. The geometrization of the gauge theories is where we are aiming at.

\section{Appendix I}
\label{appendixI}

We will discuss in this first appendix the parametrization of the local $SU(3)$ elements that represent the quotient $SU(3) / SU(2)$. We start by building the local $SU(3)$ group element objects $S = \exp \{ (\imath/2)\:\sum_{i=4}^{8} \theta_{i}\:X_{i} \}$. The $\theta_{i}, i=4\cdots8$ are all local scalars. The $X_{i}, i=4\cdots8$ are the remaining five $SU(3)$ group generators, other than the three in the local $SU(2)$ subalgebra. If the exponential matrix is fully calculated, from the condition that $\det S=1$ it emerges that there is an isomorphism between $SU(3) / SU(2)$ and $S^{5}$. We will take advantage of this isomorphism to reparameterize the original quotient elements $S$. Let us introduce the coordinates of the stereographic projections for the unit sphere $S^{5}$. The local 5-sphere is defined through $\sum_{i=1}^{6} x_{i}^{2} = 1$ where the $x_{i},\: i=1 \cdots 6$ are local coordinates. Following closely chapter III in \cite{CBDW} an in order to construct an atlas we let $P$ and $Q$ be the north and south poles respectively. Let $U=S^{5}-{P}$ and $V=S^{5}-{Q}$, let g and h be the stereographic projections of the poles $P$ and $Q$ on the plane $x_{6}=0$,

\begin{center}
$\: g: U \rightarrow \Re^{5} \:\:\: by  \:\:\:  y_{i} = x_{i} / (1-x_{6}) \:\:\: for  \:\:\:  i=1 \cdots 5$
\end{center}

\begin{center}
$\: h: V \rightarrow \Re^{5} \:\:\: by  \:\:\: z_{i} = x_{i} / (1+x_{6}) \:\:\: for  \:\:\: i=1 \cdots 5$
\end{center}

See \cite{CBDW} for the proof that it is an atlas. Then, knowing that the coordinates are local because they are defined at every point in spacetime, we can replace the local scalars $\theta_{i}, i=1 \cdots 5$ for the local atlas coordinates. We might ask what is the advantage of this procedure that feedsback new parameters for the local coset elements. We believe that the local coset representatives parameterized in terms of the local stereographic projections present a clear relationship between the coset elements and the local unit sphere $S^{5}$. It is also evident that when we calculate fully the matrix for $S = \exp \{ (\imath/2)\:\sum_{i=4}^{8} y_{i}\:X_{i} \}$ or $S = \exp \{ (\imath/2)\:\sum_{i=4}^{8} z_{i}\:X_{i} \}$ we will get again for the condition $\det S=1$ a new equation that signals the isomorphism to $S^{5}$ but now in new coordinates. We simply reparameterized the local coset elements in terms of the local stereographic projections atlas coordinates in order to establish a clear relationship between the quotient space $SU(3) / SU(2)$, that is the unit 5-sphere, and the coset representatives. For a further advantage of this coset parametrization see section \ref{appendixII}.

\section{Appendix II}
\label{appendixII}

Let us call $\exp \left\{X\right\} = \exp \{ (\imath/2)\:\sum_{i=1}^{3} \theta_{i}\:X_{i} \}$ and $\exp \left\{Y\right\} = \exp \{ (\imath/2)\:\sum_{j=4}^{8} \Psi_{j}\:X_{j} \}$. The three $\theta_{i}, i=1\cdots3$ and five $\Psi_{j}, j=4\cdots8$ are all local scalars. The first object $\exp\left\{X\right\}$ belongs to the local subgroup and the second one $\exp\left\{Y\right\}$ is a local coset representative. We name the generators following the same order as in reference \cite{GMS}. Then, $X_{8}$ is the diagonal one $\left(\frac{1}{\sqrt{3}}, \frac{1}{\sqrt{3}}, \frac{-2}{\sqrt{3}}\right)$. We would like to study the object $\exp{\left\{X\right\}}\:\exp{\left\{Y\right\}}\:\exp{\left\{-X\right\}}$. First we expand the exponential $\exp{\left\{Y\right\}} = \textbf{1} + Y + \frac{1}{2!}\:Y^{2} + \cdots$. Then we observe the following,

\begin{eqnarray}
\exp{\left\{X\right\}}\:\exp{\left\{Y\right\}}\:\exp{\left\{-X\right\}} = \exp{\left\{X\right\}}\:\left( \sum_{n=0}^{\infty} \frac{1}{n!}\:Y^{n} \right)\:\exp{\left\{-X\right\}} \nonumber \\
= \sum_{n=0}^{\infty} \frac{1}{n!}\:\exp{\left\{X\right\}}\:\left(Y^{n}\right)\:\exp{\left\{-X\right\}} =  \sum_{n=0}^{\infty} \frac{1}{n!}\:\left(\exp{\left\{X\right\}}\:Y\:\exp{\left\{-X\right\}}\right)^{n} \nonumber \\
= \exp \left( \exp{\left\{X\right\}}\:Y\:\exp{\left\{-X\right\}}\right) .\label{prevhad}
\end{eqnarray}

Now we proceed to apply the Hadamard formula \cite{RGLG} to the exponent in (\ref{prevhad}),

\begin{eqnarray}
\exp{\left\{X\right\}}\:Y\:\exp{\left\{-X\right\}} = Y + \left[X , Y\right] + \frac{1}{2!}\:\left[X , \left[X , Y\right]\right] + \cdots \label{hadamard}
\end{eqnarray}

Next we evaluate the commutator $\left[X , Y\right]$,

\begin{eqnarray}
\left[X , Y\right] &=& \left[ (\imath/2)\:\sum_{i=1}^{3} \theta_{i}\:X_{i}\: , \:(\imath/2)\:\sum_{j=4}^{8} \Psi_{j}\:X_{j} \right]
= (\imath/2)^{2}\:\sum_{i=1}^{3}\:\sum_{j=4}^{8} \theta_{i}\:\Psi_{j}\:\left[X_{i}\: ,\: X_{j}\right] \nonumber \\
&=&  (\imath/2)^{2}\:\sum_{i=1}^{3}\:\sum_{j=4}^{8} \theta_{i}\:\Psi_{j}\:(2\:\imath)\:\sum_{k=4}^{7}\:f_{ijk}\:X_{k}
= (\imath/2)\:\sum_{k=4}^{7}\:\left\{ - \sum_{i=1}^{3}\:\sum_{j=4}^{8} \theta_{i}\:\Psi_{j}\:f_{ijk} \right\}\:X_{k}  \nonumber \\
&=& (\imath/2)\:\sum_{k=4}^{7}\:n_{k}\:X_{k} \ . \label{firsthadamard}
\end{eqnarray}

Observe that the relation $\left[X_{i}\: ,\: X_{j}\right] = (2\:\imath)\:\sum_{k=4}^{7}\:f_{ijk}\:X_{k}$ for $i:1\ldots3, j:4\ldots8$ is key in our deduction. Likewise, the next commutator in the Hadamard expansion yields,

\begin{eqnarray}
\left[ X , \left[X , Y\right]\right] &=& \left[ (\imath/2)\:\sum_{i=1}^{3} \theta_{i}\:X_{i}\: , \:(\imath/2)\:\sum_{k=4}^{7} n_{k}\:X_{k} \right] \nonumber \\
&=& (\imath/2)\:\sum_{h=4}^{7}\:m_{h}\:X_{h} \ , \label{secondhadamard}
\end{eqnarray}

for some $m_{h},\: h:4\ldots7$. Therefore, when we sum the Hadamard expansion we obtain,

\begin{eqnarray}
\exp{\left\{X\right\}}\:Y\:\exp{\left\{-X\right\}} = (\imath/2)\:\sum_{p=4}^{8}\:\phi_{p}\:X_{p} \ ,  \label{sumhadamard}
\end{eqnarray}

for some $\phi_{p},\: p:4\ldots8$. Now, we can call,

\begin{eqnarray}
A = \sum_{p=4}^{8}\:\phi_{p}^{2} = \frac{\left(1 - x_{6}^{2}\right)}{\left(1 \pm x_{6}\right) }  \ .  \label{sixthcoord}
\end{eqnarray}

where the $\pm$ refers to the two posible charts in our parametrization, see section \ref{appendixI}. Finally we find,

\begin{eqnarray}
x_{6} = \frac{\pm A \mp 1 }{ A + 1}  \ .  \label{sixthcoordfinal}
\end{eqnarray}

We conclude that the original object $\exp{\left\{X\right\}}\:\exp{\left\{Y\right\}}\:\exp{\left\{-X\right\}}$ is a coset representative. When we analyze the memory of the transformation in equation (\ref{S1DOUBLETR}) we can see that the object $S_{ROT1}\:S_{2}\:S_{ROT1}^{-1}$ is a coset element as well. Therefore, there is no memory in the subsequent transformations.

\section{Appendix III}
\label{appendixIII}

The mathematical ideas developed in appendix I in paper \cite{A2} can be adapted to our present $SU(3)$ case. In paper \cite{A2} the derivative $\partial_{\tau}\:(S)$ was studied for a local $SU(2)$ group element $S$. Through this study we concluded and we quote ``The $SU(2)$ group of local gauge transformations, generates proper and improper LB1 transformations. Therefore the image of $SU(2)$ is not associated to a subgroup of LB1 (or tensor products of LB1)''. In this paper we have to consider instead for a similar analysis the derivative $\partial_{\tau}\:(S\:S_{ROT})$. Explicitly we can write $\partial_{\tau}(S)\:S_{ROT} + S\:\partial_{\tau}(S_{ROT})$. Therefore similar conclusions can be drawn from the analysis of the factor $\partial_{\tau}(S_{ROT})$ since $S_{ROT}$ is an element in the local $SU(2)$ subalgebra. The addition of the first term $\partial_{\tau}(S)\:S_{ROT}$ or the local coset factor $S$ in the second term is not going to alter the summary in appendix I of paper \cite{A2}. Simply because we can make the factor $\partial_{\tau}(S_{ROT})$ as large positive or negative as we wish, therefore making the whole object $\partial_{\tau}(S)\:S_{ROT} + S\:\partial_{\tau}(S_{ROT})$ as large positive or negative as we wish in all its components. Summarizing, we will reach the same conclusions in the local $SU(3)$ case as we reached in paper \cite{A2} for the local $SU(2)$ case. Again we want to prove that the image of $SU(3)$ is not associated to a subgroup of LB1 (or tensor products of LB1). In order to illustrate this last assertion we will analyze a particular example. This example will be studied in all three cases Abelian $U(1)$, non-Abelian $SU(2) \times U(1)$ and non-Abelian $SU(3) \times SU(2) \times U(1)$.

\subsection{Abelian $U(1)$}
\label{exampleu1}

We will make use in this Abelian example, of the results, expressions and analysis done in the section Gauge Geometry in paper \cite{A}. Let us remember first, and we will transcribe a few important expressions from this last section in order for our example to be easily followed. First the local scalars that define the nature of the local tetrad transformation induced by a local Abelian gauge transformation associated to the local scalar $\Lambda$. Let us also remember that the tetrad vectors $V_{(1)}^{\beta}$ and $V_{(2)}^{\beta}$ generate at every point in spacetime blade one, and that they are not normalized.

\begin{eqnarray}
C&=&(-Q/2)\:V_{(1)\sigma}\:\Lambda^{\sigma} / (\:V_{(2)\beta}\:
V_{(2)}^{\beta}\:)\label{COEFFC}\\
D&=&(-Q/2)\:V_{(2)\sigma}\:\Lambda^{\sigma} / (\:V_{(1)\beta}\:
V_{(1)}^{\beta}\:)\ .\label{COEFFD}
\end{eqnarray}

We would like to calculate the norm of the transformed vectors
$\tilde{V}_{(1)}^{\alpha}$ and $\tilde{V}_{(2)}^{\alpha}$,

\begin{eqnarray}
\tilde{V}_{(1)}^{\alpha}\:\tilde{V}_{(1)\alpha} &=&
[(1+C)^2-D^2]\:V_{(1)}^{\alpha}\:V_{(1)\alpha}\label{FP}\\
\tilde{V}_{(2)}^{\alpha}\:\tilde{V}_{(2)\alpha} &=&
[(1+C)^2-D^2]\:V_{(2)}^{\alpha}\:V_{(2)\alpha}\ ,\label{SP}
\end{eqnarray}

where the relation $V_{(1)}^{\alpha}\:V_{(1)\alpha}=
-V_{(2)}^{\alpha}\:V_{(2)\alpha}$ has been used.

Then, let us proceed to our particular case where we consider the following situation at some point in spacetime. $1 + C > D > 0$, where $0 > C > -1$. This last case represents a locally proper tetrad transformation, a boost. Next, let us consider a new gauge transformation at the same point, $A\:\Lambda$, where $A$ is a constant. Then, the new $C_{new} = A\: C$ and $D_{new} = A\: D$ characterize the new tetrad transformation on blade one. In addition let this constant $A$ be such that $1 - A\:\mid C \mid = \epsilon$, $0 < \epsilon \ll 1$. The idea behind all this introduction would be to turn the original boost tetrad transformation into an improper transformation. For instance an improper transformation that satisfies $A\:D = D_{new} > 1 + C_{new} = 1 - A\:\mid C \mid = \epsilon$. This last improper condition is equivalent, after some algebra, to demanding that $\epsilon < 1 / (1 + \mid C \mid / D)$. Knowing that $A = (1 - \epsilon) / \mid C \mid$, by choosing at the point under consideration a constant $\epsilon$ that satisfies the inequality given above, $\epsilon < 1 / (1 + \mid C \mid / D)$, we would be generating an improper tetrad transformation , $A\:\Lambda$, out of a proper boost $\Lambda$. The meaning of this result is nothing but stating that the image of the local electromagnetic gauge transformations into local tetrad transformations is not a subgroup of the $LB1$ group. Simply because we proved that a local electromagnetic gauge transformation $\Lambda$ inducing a proper local Lorentz tetrad transformation, can be turned into an improper tetrad Lorentz transformation through the new local electromagnetic gauge transformation $A\:\Lambda$. This last remark is equivalent to say that this mapping developed in paper \cite{A} is surjective. The image is LB1, and not one of its subgroups.

\subsection{Non-Abelian $SU(2) \times U(1)$}
\label{examplesu2}

In the non-Abelian case we would be repeating a similar line of arguments as those exposed in the Abelian case. Nonetheless there will be particular remarks in this non-Abelian situation that we would like to make. This subject was thoroughly discussed in appendix I in paper \cite{A2}, we will transcribe here the minimum number of elements necessary in order to make our presentation consistent, therefore the notation will be the same as in paper \cite{A2}. In the first place the new local scalars and tetrads are given (see section gauge geometry in \cite{A2}) by the following expressions,

\begin{eqnarray}
C^{'} &=&{\imath \over g}\: (-Q_{ym}/2)\:\:S^{'\sigma\:}_{(1)}\: Tr[\tilde{\Lambda}^{\alpha}_{\:\:\:\delta}\:\tilde{\Lambda}^{\beta}_{\:\:\:\gamma}\:\Sigma^{\delta\gamma}\:E_{\alpha}^{\:\:\rho}\: E_{\beta}^{\:\:\lambda}\:\ast \xi_{\rho\sigma}\:\ast \xi_{\lambda\tau}\:\partial^{\tau}\:(S)\:S^{-1}] / (\:S^{'}_{(2)\mu}\:S_{(2)}^{'\mu}\:) \label{COEFFCPRIMA}\\
D^{'} &=&{\imath \over g}\: (-Q_{ym}/2)\:\:S^{'\sigma\:}_{(2)}\: Tr[\tilde{\Lambda}^{\alpha}_{\:\:\:\delta}\:\tilde{\Lambda}^{\beta}_{\:\:\:\gamma}\:\Sigma^{\delta\gamma}\:E_{\alpha}^{\:\:\rho}\: E_{\beta}^{\:\:\lambda}\:\ast \xi_{\rho\sigma}\:\ast \xi_{\lambda\tau}\:\partial^{\tau}\:(S)\:S^{-1}] / (\:S^{'}_{(1)\mu}\:S_{(1)}^{'\mu}\:) \ . \label{COEFFDPRIMA}
\end{eqnarray}

We would like as well, to calculate the norm of the transformed vectors
$\tilde{S}_{(1)}^{\mu}$ and $\tilde{S}_{(2)}^{\mu}$,

\begin{eqnarray}
\tilde{S}_{(1)}^{\mu}\:\tilde{S}_{(1)\mu} &=&
[(1+C^{'})^{2}-D^{'2}]\:S_{(1)}^{'\mu}\:S^{'}_{(1)\mu}\label{FP}\\
\tilde{S}_{(2)}^{\mu}\:\tilde{S}_{(2)\mu} &=&
[(1+C^{'})^{2}-D^{'2}]\:S_{(2)}^{'\mu}\:S^{'}_{(2)\mu}\ ,\label{SP}
\end{eqnarray}

where the relation $S_{(1)}^{'\mu}\:S^{'}_{(1)\mu}=
-S_{(2)}^{'\mu}\:S^{'}_{(2)\mu}$ has been used.

We write the elements in $SU(2)$ as,

\begin{eqnarray}
S = \sigma_{o}\:\cos(\theta/2) + \imath \: \sigma_{j}\:\hat{\theta}^{j}\:\sin(\theta/2) = \sum\:{1 \over n!}
\left( \sum_{i=1}^{3} {\imath \over 2 } \sigma_{i}\: \theta^{i}\right)^{n}\ ,\label{S}
\end{eqnarray}

\begin{equation}
\theta^2 = \sum_{i=1}^{3}\:\left( \theta^{i}\right)^2\ ,\:\:\:\:\: \mbox{where $\hat{\theta}^{i}$ is given by,} \:\:\:\:\:\hat{\theta}^{i} = \theta^{i} / \mid \theta \mid \ .\label{thetahat}
\end{equation}

$\sigma_{o}$ is the identity, $\sigma_{j}$ for $j=1\ldots3$ are the usual Pauli matrices, and the summation convention is applied for $j=1\ldots3$. We can write the derivative $\partial_{\lambda}S$ where $S$ is a local $SU(2)$ gauge transformation as,

\begin{equation}
\partial_{\lambda}S_{\mid_{\theta=0}} = \left[(-1/2)\:\sigma_{o}\:\sin(\theta/2)\:\partial_{\lambda}\theta + (\imath/2)\:\sigma_{j}\hat{\theta}^{j}\:\cos(\theta/2)\:\partial_{\lambda}\theta + \imath\:\sigma_{j}\:\partial_{\lambda}\hat{\theta}^{j}\:\sin(\theta/2)\right] _{\mid_{\theta=0}} \ .\label{partialS}
\end{equation}

Again as in paper \cite{A2} we argue that since the vector components of $Tr[\tilde{\Lambda}^{\alpha}_{\:\:\:\delta}\:\tilde{\Lambda}^{\beta}_{\:\:\:\gamma}\:\Sigma^{\delta\gamma}\:E_{\alpha}^{\:\:\rho}\: E_{\beta}^{\:\:\lambda}\:\ast \xi_{\rho\sigma}\:\ast \xi_{\lambda\tau}\:\partial^{\tau}\:(S)\:S^{-1}]$, can take on any values, positive or negative, then we must conclude that $1+C^{'}$ and $D^{'}$ can take on any possible real values. Borrowing again the ideas from \cite{A}, and section \ref{exampleu1} we can analyze as an example, the case where $1+C^{'} > D^{'} > 0$, and $0 > C^{'} > -1$. Let us suppose in addition that $\partial_{\rho}\theta$, $\partial_{\rho}\hat{\theta}^{i}$ and $\hat{\theta}^{i}$ have finite components at the origin. We can always consider the geodesic through the origin of the $2\pi$ sphere, such that $\theta^{i}_{A} = A\: \theta^{i}$, where $A$ is a constant. Now, $\theta_{A} = A\:\theta$ and $\hat{\theta}^{i}_{A} = \hat{\theta}^{i}$, but $\partial_{\rho}\theta_{A} = A\:\partial_{\rho}\theta$. Then, at the origin of the parameter sphere, $\theta^{i} = 0$, $\theta_{A} = 0$ and $\partial_{\rho}\theta_{A\mid\theta = 0} = A\:\partial_{\rho}\theta\mid_{\theta=0}$.  Then, the new $C^{'}_{new} = A\: C^{'}$ and $D^{'}_{new} = A\: D^{'}$ characterize the new tetrad transformation on blade one. If we proceed exactly as in the Abelian case presented in the subsection \ref{exampleu1} above, we will reach exactly the same conclusion. The $SU(2)$ group of local gauge transformations, generates proper and improper LB1 transformations. Therefore the image of $SU(2)$ is not associated to a subgroup of LB1 (or tensor products of subgroups of LB1).

\subsection{Non-Abelian $SU(3) \times SU(2) \times U(1)$}
\label{examplesu3}

Let us import all the ideas from the two subsections above and we will have all necessary elements to analyze the non-Abelian $SU(3) \times SU(2) \times U(1)$ case. But we just need expression (\ref{S1L2})

\begin{eqnarray}
Tr[\tilde{\Lambda}^{\alpha}_{\:\:\:\delta}\:\tilde{\Lambda}^{\beta}_{\:\:\:\gamma}\:\Sigma_{5}^{\delta\gamma}\:S_{\alpha}^{\:\:\rho}\: S_{\beta}^{\:\:\lambda}\:\ast \epsilon_{\rho\sigma}\:\ast \epsilon_{\lambda\tau}\:A^{\tau}] + \nonumber \\ {\imath \over g} \: Tr[\tilde{\Lambda}^{\alpha}_{\:\:\:\delta}\:\tilde{\Lambda}^{\beta}_{\:\:\:\gamma}\:\Sigma_{5}^{\delta\gamma}\:S_{\alpha}^{\:\:\rho}\: S_{\beta}^{\:\:\lambda}\:\ast \epsilon_{\rho\sigma}\:\ast \epsilon_{\lambda\tau}\:\partial^{\tau}\:(S\:S_{ROT})\:(S\:S_{ROT})^{-1}] \ .
\label{S1L2app}
\end{eqnarray}

We can rewrite the expression above in terms of $\partial_{\tau}(S)\:S_{ROT} + S\:\partial_{\tau}(S_{ROT})$. If we follow all the steps in subsection gauge geometry in paper \cite{A2} and subsection \ref{examplesu2} we can get to the point where again a similar problem to the one in the two previous subsections is posed to us. Now, paying attention to expressions (\ref{COEFFCPRIMA}-\ref{COEFFDPRIMA}), we can see that $C^{'}_{new} = C^{'}_{coset} + A\: C^{'}_{su(2)}$ and $D^{'}_{new} = D^{'}_{coset} + A\: D^{'}_{su(2)}$ characterize the new tetrad transformation on blade one. $C^{'}_{coset}$ will have its origin in the term  $\partial_{\tau}(S)\:S_{ROT}$ while $C^{'}_{su(2)}$ will have its origin in the term $S\:\partial_{\tau}(S_{ROT})$.  Again, we can analyze as an example, the case where $1+C^{'} > D^{'} > 0$, and $0 > C^{'} > -1$. We are also assuming as a particular example that locally at the point in consideration the following conditions are satisfied, $0 > C^{'}_{coset} > -1$, $0 > C^{'}_{su(2)} > -1$, $D^{'}_{coset} > 0$ and $D^{'}_{su(2)} > 0$. If we consider a constant $A$ such that $1 - \mid C^{'}_{coset} \mid - A\:\mid C^{'}_{su(2)} \mid = \epsilon$, $0 < \epsilon \ll 1$, and we also want this new gauge transformation to fulfill $D^{'}_{coset} + A\:D^{'}_{su(2)}  = D^{'}_{new} > 1 + C^{'}_{new}  = 1 - \mid C^{'}_{coset} \mid - A\:\mid C^{'}_{su(2)} \mid = \epsilon$, then the following inequality will have to be satisfied by $\epsilon$,

\begin{equation}
\epsilon\:\left( (1/D^{'}_{su(2)}) + (1/\mid C^{'}_{su(2)} \mid) \right)  < \left( ((1 - C^{'}_{coset}) / \mid C^{'}_{su(2)} \mid) + (D^{'}_{coset}/D^{'}_{su(2)}) \right) \ . \label{epcondsu3}
\end{equation}

By choosing at the point under consideration a constant $\epsilon$ to satisfy inequality (\ref{epcondsu3}) then we would be turning an original proper boost into an improper tetrad transformation. The meaning of this result is nothing but stating that the image of the local $SU(3)$ transformations into local tetrad transformations is not a subgroup of the $LB1$ group (or tensor products of subgroups of LB1). This last remark is equivalent to say that this mapping is surjective.


\begin{thebibliography}{99}


\bibitem{A} A. Garat, J. Math. Phys. {\bf 46}, 102502 (2005). A. Garat, Erratum: Tetrads in geometrodynamics, J. Math. Phys. {\bf 55}, 019902 (2014).

\bibitem{A2} A. Garat, Tetrads in Yang-Mills geometrodynamics, Gravitation and Cosmology, (2014) Vol. {\bf 20} No. 1, pp. 116-126. Pleiades Publishing Ltd. arXiv:gr-qc/0602049.

\bibitem{A3} A. Garat, The new electromagnetic tetrads, infinite tetrad nesting and the non-trivial emergence of complex numbers in real theories of gravitation, Int. J. Geom. Methods Mod. Phys., Vol. {\bf 14}, No. 9 (2017), 1750132.

\bibitem{MW} C. Misner and J. A. Wheeler, Annals of Physics {\bf 2}, 525 (1957).

\bibitem{RG} R. Geroch, {\it General Relativity: From A to B} {(University of Chicago Press, Chicago, 1978)}.

\bibitem{RW} R. ~Wald, {\it General Relativity} (University of Chicago Press, Chicago, 1984).

\bibitem{MC2} M. ~Carmeli, {\it Classical Fields: General Relativity and Gauge Theory} (J. Wiley \& Sons, New York, 1982).

\bibitem{EJKS} J. Ehlers, P. Jordan, W. Kundt and R. Sachs, Akad. Wiss. Lit. Mainz Abh. Math.-Natur. Kl, {\bf 11} (Mainz 4), 793 (1961).

\bibitem{KK} K. Kucha\v{r}, Phys. Rev D {\bf 4}, 955 (1971); J. Math. Phys. {\bf 13}, 768 (1972); {\bf 17}, 777 (1976); {\bf 17}, 792 (1976); {\bf 17}, 801 (1976); {\bf 18}, 1589 (1977).

\bibitem{JDBKK} J. D. ~Brown and K. ~Kucha\v{r}, Phys. Rev D, {\bf 51}, 5600 (1995).

\bibitem{FM} A. E. Fischer and J. E. Marsden, J. Math. Phys. {\bf 13}, 546 (1972).

\bibitem{FAEP} F. A. E. Pirani, {\it Les Theories Relativistes de la Gravitation} (CNRS, Paris, 1962).

\bibitem{JWY1} J. W. ~York, J. Math. Phys. {\bf 13}, 125 (1972); {\bf 14}, 456 (1973)

\bibitem{JWY2} J. W. ~York, Phys. Rev D, {\bf 10}, 428 (1974).

\bibitem{NOMJWY} N. ~O`Murchadha and J. W. ~York, {\bf 14}, 1551 (1973).

\bibitem{HY} H. P. Pfeiffer and J. W. York, Phys. Rev D, {\bf 67}, 044022 (2003).

\bibitem{JY} R. T. Jantzen and J. W. York /gr-qc 0603069 (2006).

\bibitem{LICH} A. Lichnerowicz, J. Math. Pure and Appl. {\bf 23}, 37 (1944).

\bibitem{YCB} Y. Choquet-Bruhat in {\it Gravitation: An Introduction to Current Research} edited by L. Witten (Wiley, New York, 1962).

\bibitem{CMDWYCB} C. M. DeWitt and Y. Choquet-Bruhat {\it Analysis, Manifolds and Physics} edited by  (North-holland, The Netherlands , 1982).

\bibitem{AEO} A. Garat, Euler observers in geometrodynamics, Int. J. Geom. Meth. Mod. Phys., Vol. {\bf 11}, No. 6 (2014), 1450060. arXiv:gr-qc/1306.4005

\bibitem{ADM} R. Arnowitt, S. Deser and C. W. Misner, {\it ``The Dynamics of General Relativity''} in {\it Gravitation: An Introduction to Current Research} edited by L. Witten (Wiley, New York, 1962).

\bibitem{EB} F. B. Estabrook, Phys. Rev. D {\bf 71}, 044004 (2005).

\bibitem{ERW} F. B. Estabrook, R. S. Robinson, H. D. Wahlquist, Class. Quant. Grav. {\bf 14}, 1237 (1997).

\bibitem{BB} L. T. Buchman, J. M. Bardeen, Phys. Rev, D {\bf 67}, 084017 (2003).

\bibitem{LSJWY} L. Smarr and J. W. York, Phys. Rev. D {\bf 17}, 2529 (1978).

\bibitem{SWNG} S. Weinberg, Phys. Rev. {\bf 139}, B597 (1965).

\bibitem{LORNG} L. O'Raifeartagh, Phys. Rev. {\bf 139}, B1052 (1965).

\bibitem{CMNG} S. Coleman and J. Mandula, Phys. Rev. {\bf 159}, N5 1251 (1967).

\bibitem{RGLG} R. Gilmore, {\it Lie Groups, Physics and Geometry} (Cambridge University Press, 2008).

\bibitem{MN} M. Nakahara, {\it Geometry, Topology and Physics} (IOP Publishing, 1990).

\bibitem{CI} C. J. Isham, {\it Modern Differential Geometry for Physicists} (World Scientific Publishing, 1989).

\bibitem{GMS} W.~Greiner and B.~Mueller, {\it Quantum Mechanics, Symmetries}
(Springer Verlag, 1989).

\bibitem{CBDEL} S. Capozziello, G. Basini and M. De Laurentis, Eur. Phys. J. {\bf C71}, 1679 (2011).


\bibitem{DESISH} S. Deser and C. J. Isham, Phys. Rev D, {\bf 14}, 2505 (1976).

\bibitem{DEHETEI} S. Deser, M. Henneaux and C. Teitelboim, Phys. Rev D, {\bf 55}, 826 (1997).

\bibitem{HETEI} M. Henneaux and C. Teitelboim, Phys. Rev D, {\bf 71}, 024018  (2005).

\bibitem{DEGOHETEI} S. Deser, A. Gomberoff, M. Henneaux and C. Teitelboim, Phys. Lett. B, {\bf 400}, 80  (1997).

\bibitem{CDEL} S. Capozziello and M. De Laurentis, Int. J. Geom. Meth. Mod. Phys., {\bf 11}, (2014) 1460004.

\bibitem{RJ} R. Jackiw, {\it Fifty Years of Yang-Mills Theory and my Contribution to it} (arXiv:physics/0403109,
 2004).

\bibitem{CBDW}  Y. Choquet-Bruhat and C. DeWitt-Morette, {\it Analysis, Manifolds and Physics}
(Elsevier Science Publishers B.V., 1987).










\end{thebibliography}

\end{document}